\documentstyle[12pt]{article}

\begin{document}

\def\be{\begin{equation}}
\def\ee{\end{equation}}
\def\bea{\begin{eqnarray}}
\def\eea{\end{eqnarray}}
\def\nn{\nonumber}
\def\ep{\epsilon}
\def\c{\cite}
\def\m{\mu}
\def\ga{\gamma}
\def\lan{\langle}
\def\ran{\rangle}

\def\Ga{\Gamma}
\def\la{\lambda}
\def\si{\sigma}
\def\al{\alpha}
\def\pa{\partial}
\def\de{\delta}
\def\De{\Delta}
\def\rsr{{r_{s}\over r}}
\def\rrs{{r\over r_{s}}}
\def\rs2r{{r_{s}\over 2r}}
\def\l2r2{{l^{2}\over r^{2}}}
\def\rsa{{r_{s}\over a}}
\def\rsb{{r_{s}\over b}}
\def\rsro{{r_{s}\over r_{o}}}
\def\rss{r_{s}}
\def\a2{{l^{2}\over a^{2}}}
\def\b2{{l^{2}\over b^{2}}}
\def\op{\oplus}
\def\sn{\stackrel{\circ}{n}}
\def\c{\cite}

\def\be{\begin{equation}}
\def\ee{\end{equation}}

\def\bea{\begin{eqnarray}}
\def\eea{\end{eqnarray}}

\begin{titlepage}
\vspace*{10mm}
\begin{center} {\Large \bf
On  the  Mass Neutrino Phase  along the geodesic  line
and the null line in curved and flat spacetime}

\vskip 10mm
\centerline {\bf C.M. Zhang$^{1,2}$,  and A.Beesham$^3$}
\vskip 1cm
{\it
1.   Research Centre for Theoretical Astrophysics, 
       School  of  Physics, \\University of Sydney, NSW 2006, Australia\\
 zhangcm@physics.usyd.edu.au\\

2.  High Energy Section, ICTP, strada Costiera 11, 
        34100 Trieste,  Italy\\
3. Department of Mathematical Sciences, 
 University of Zululand,   South Africa}

\end{center}

\vskip 0.5cm

\begin{abstract}
\noindent
On  the mass neutrino phase calculations along both the particle   
geodesic line  and the photon null line, 
there exists a double counting, factor of 2,  when comparing 
the geodesic phase with the null phase. 
 Moreover,  we compare the phase calculations 
among the same energy description, 
the same momentum description 
by means of the Minkowski diagram, and obtain the practical equivalence 
of these two descriptions. On the same velocity description, although it 
does not correspond to a reality of physical process, we still indicate 
its phase calculation in the Minkowski diagram,  which has the same 
result as those of same energy and same momentum cases. 
 Further, in the curved spacetime, we also 
prove the existence of the double counting of the geodesic phase to the
null phase. Our conclusions are same as others' results by 
 the different methods.  

\end{abstract}

\end{titlepage}

\newpage

\section{Introduction}

There have been many topics issued  on the quantum mechanics of the
neutrino oscillation in the flat
spacetime ~\cite{zub98,bgg98,pak98,bp78,boe92} and in the curved
spacetime 
\cite{ful96,for96,koj96,gro96,bha99,zcm20,zcm01,zcm21}, most of them
centered on
the understanding of the quantum coherence condition between the various
mass-eigenstates, as well as on the calculations  of the phase
factor in the spacetime. The mass neutrino
oscillation problem is connected basically to the phase difference between
different mass-eigenstate, 
a property intimately related to the basic principles
of quantum mechanics. In the standard treatment of the neutrino oscillation, the
condition of same-energy (same-momentum) with different momenta (energies) of the
mass-eigenstates is introduced~\cite{bgg98,bp78,boe92,dod97,oku00,tsu01}, 
 which arises the same practically applied massive neutrino oscillation phase, 
however it is indicated that the unambiguous theoretical description of the 
massive neutrino oscillation phase will be involved in the wave packet formalism 
and not in the plane wave approximation \c{dod99}. Further, it is shown that 
the same energy or same momentum description 
 is just the arbitrariness of the choice of the Lorentz 
frame in which it is valid,  nothing to do with the physical argument \c{giu01}. 
The change of flavors is described by the oscillation length, and it
is usually claimed that both conditions (same-energy and same-momentum) present
practically the same neutrino oscillation results. Sometimes, a
source-dependent condition is added~\cite{kay81,win82,gol96}, which
implies
in giving up either the same-energy or the same-momentum prescription. This
calculation of the neutrino oscillation phase, however, yields the same result
of the standard treatment. 
Furthermore, the velocity difference of the various mass-eigenstates results in
a spacetime separation for neutrinos of different
flavor~\cite{gro96,lip95,nie96}, which is another source of confusion on the
interference condition for the different neutrino. 
The same velocity description has been paid much attention \c{tak98,leo99,tak00}, 
 but this prescription is pointed out to be forbidden kinematically \c{oku007}.  
Actually, considerable
confusion has arisen in the description, interpretation and understanding of the
neutrino oscillation, a problem which involves the fundamental principles of both
quantum mechanics and  special relativity, such as the uncertainty principle, the
superposition principle, and simultaneity problems because the mass neutrinos are 
high energy quantum objects. 
 Moreover it is often noted the factor of 2  of the neutrino phase
calculations in the flat spacetime
\cite{lip95,lip98} and in the curved spacetime
\cite{gro96,bha99}, which is believed  to be the
consideration of the 
space  phase or the time phase \cite{gro96}, as well as the arrival time
difference of
the two mass neutrinos \cite{bha99}.

In the Minkowski diagram of flat spacetime, we discuss how  the 
same-energy,  the same-momentum  of the 
neutrino phase present the same practical result, and 
the relation between the arrival time difference and   the double counting 
of the phase,  and further, we extend our
discussions  to the curved spacetime. We also calculate the phase factor 
 in the case of the same velocity description although this process is not 
 physically realized. 
 Here we stress that all our calculations are based on the
plane wave treatment of the neutrino, otherwise the wave packet should be considered 
\c{giu91}. We set $\hbar = c = 1$ throughout this article.

\section{The Standard Treatment of the Neutrino Oscillation}
        
In the standard treatment of the neutrino oscillation, if two generations
are taken into account, the neutrino flavor-state is written as 
\be
|\nu _{\alpha}(x,t) \rangle = \sum_j U_{\alpha j} 
\exp \left[ i\Phi_j \right] | \nu _j \rangle \; ,  
\ee 
where
\be
(U) = U_{\alpha j}= \left( \begin{array}{cc}
\cos \theta & \sin \theta \\
 -\sin \theta & \cos \theta
\end{array} \right) \; ,
\ee            
with $\theta$ the mixing angle. In vector form,
\be
\left( \begin{array}{c}
|\nu_{e}(x,t) \rangle\\
|\nu_{\mu}(x,t) \rangle
\end{array} \right)
= \exp(i \bf{\Phi })~U
\left( \begin{array}{c}
|\nu _{1} \rangle\\
|\nu _{2} \rangle
\end{array} \right) \; ,
\ee
where   
\be
\left( \exp(i \bf{\Phi })\right)= \left( \begin{array}{cc}
\exp(i \Phi _{1}) & 0 \\
0 & \exp(i \Phi _2)
\end{array} \right),
\ee
with $\Phi_j$ the eigenvalue of the phase operator~\cite{ful96}
\be
\Phi_j = \int \left( E_{j} \, dt - P_{j} \, dx \right) \; .
\ee
Here, flavor (mass) indices are expressed by Greek (Latin) letters, and
$\nu_{e}$ and $\nu_{\mu}$ are represented respectively by 
$\nu_{1}$ and $\nu_{2}$. The matrix elements $U_{\alpha j}$ comprise the
transformation between the flavor and mass basis. Now, we suppose a pure
flavor-state electron  neutrino $| \nu _e\rangle$ is at the initial source 
position $\bf{A}$ ($x=0$ and  $t=0$). The mass eigenstates are taken to be the
eigenstates of energy (momentum).  The momentum (energy) then satisfies the mass
shell relation 
\be
{\bf P} = \sqrt{{\bf E}^2 - {\bf m}^2} \approx {\bf E} - 
\frac{{\bf m}^2}{2 {\bf E}} \; ,
\label{pe}
\ee
where ${\bf m}$ is the rest-mass operator corresponding to
the mass eigenvalue $m_j$ of the  eigenstate $| \nu_j \rangle$~\cite{ful96}.

The oscillation probability ${\cal P}$ from flavor $|\nu_e
\rangle$ at the source  position {\bf A}, to flavor  $|\nu_{\mu}
\rangle$ at the detector position $\bf{B}$ ($x=L$) is given
by~\cite{boe92}, 
\be
{\cal P}(\nu_e \rightarrow  \nu _{\mu})
=|\langle \nu _{\mu}|\nu _e(x,t) \rangle|^2
= \sin^{2}2\theta ~\cos{\Delta \Phi \over 2},
\ee
where
\be
\Delta \Phi = \Phi _2 - \Phi _1 
\ee
is the phase difference between the different mass states. The oscillation
length is defined by taking $\Delta \Phi = 2\pi$. From the standard treatment
of the neutrino oscillation, the common momentum (energy), and the approximation
condition (\ref{pe}) lead to a phase difference $\Delta \Phi$ and
an oscillation length $L_{osc}^{S}$ given respectively by~\cite{boe92}
\be
\Delta \Phi = {\Delta m^2 \over 2E} L \; ,
\ee
and
\be
L_{osc}^{S}= 2\pi {2E \over \Delta m^2} \; ,
\ee
where $ \Delta m^2= m_2^2 -m_1^2$, with $m_2 >m_1$. To compute the oscillation
probability, the following three assumptions are often
applied~\cite{bp78}: ($i$) The mass eigenstates are taken to be the energy
(momentum) eigenstates, with a common energy (momentum); ($ii$) up to
${\cal O}$($m/E$), we have the approximation $P \approx E >> m$; ($iii$) a
massless trajectory is assumed, which means that the neutrino travels along the
null trajectory defined by $dx=dt$. With these assumptions, the flavor state is
simplified as
\be
|\nu _{\alpha}(x,t) \rangle = \sum_j U_{\alpha j} \exp \left[ -i
        \left(m_j^2 \over 2E\right) x \right] | \nu _j
         \rangle. 
\label{}
\ee
This state is then used to compute the oscillation amplitude.

\section{Reexamination of the Standard Treatments}

In this section, we discuss the standard procedures for the calculation of the
phase factor along both the particle world-line and the null trajectory in
Minkowski spacetime~\cite{fre69}. First, let us recall that, when the null
condition is applied, the standard treatment yields 
\be 
\Phi ^S = \int (Edt - Pdx) \approx  \int (E - P)dx = 
{m^2 \over 2E} L \; .      
\ee
If the null condition $dx = dt$ is not used, we find 
\be
\Phi  = \int (Edt - Pdx) =  \int {(E^2 - P^2)\over P}dx =
{m^2 \over P}L \approx 2\Phi ^S \,.
\ee 
The phase difference is then
\be
\Delta \Phi = \left({m^2_2 \over P_2} - {m^2_1 \over P_1} \right)L 
\approx 2\Delta \Phi ^S \, .
\ee

In order to illustrate the problem explicitly, the original definition 
of the phase factor will be obtained by means of the interval in the
Minkowski spacetime. From Fig.\ref{figsv}, we see that the interval of
the world
line, when the neutrino propagates from ${\bf A}$ to ${\bf B}$ in spacetime,
is 
\be
ds^2(AD) = {\overline{BD}}^2 - {\overline{BN}}^2 
= (\overline{BD} + \overline{BN})\overline{ND} 
= {L \over v}^2 - L^2 
= {L^2 \over v^2 \gamma ^2} \; ,
\label{ds}
\ee
and consequently 
\be
\Phi  = m~ ds(AD) = {mL\over v\gamma }= \left({m^2 \over
E}L\right)\left({1\over v}\right)
\approx 2\Phi ^S \,.
\ee
The standard treatment gives
\be
m~ ds(AD) = \left({m \overline{BD}\over ds}\right) \overline{BD} 
- \left({m \overline{BN}\over ds}\right) \overline{BN}
= (E - P)\overline{BN}= {m^2 \over 2E}L=\Phi^S \; .
\label{dsps}
\ee
Although the difference $\overline{ND}= \overline{BD} - \overline{BN}$ 
between the particle world time and the null time  is small, the phase  
is not solely related to the distance in space, but also to the interval 
in the spacetime, and the phase is sensitive to the null condition when the
velocity of the neutrino approaches the speed of light. If we neglect this
small difference in the time interval, a factor of 2 will appear in the
phase factor! To illustrate this conclusion, we inspect in the next
subsections the standard treatments in more detail. 

\subsection{Reexamination of the Same-Energy Prescription}

If two neutrinos share the same energy ($E_1 = E_2 = E$, 
$E^2 - P_j^2 = m_j^2$), but present different momenta, 
the contribution to the phase difference will come from the 
integration of the momentum in space because the same-energy condition 
makes the integration of the energy in time to vanish. This is the main 
point of the same-energy condition in the standard treatment. Now, we will
explore this point in more detail by using the Minkowski diagram of
Fig.\ref{figdv}. 

For convenience, we will keep using the convention 
$m_2 > m_1$, which leads to $\gamma_1 > \gamma_2$, 
$P_1 > P_2$ and $v_1 > v_2$. In the 
Minkowski diagram of Fig.\ref{figdv}, the faster the frame the closer it
is of 
the null line~\cite{fre69}. This property will be helpful for our analysis.

According to the standard treatment in the case of same-energy ($E_1 =
E_2$), we have
\bea 
\label{edps}
\Delta { \Phi}^S &=& \int_{A}^{B} (E_2 dt - P_2 dx) - 
\int_{A}^{B} (E_1 dt - P_1 dx) \nonumber \\  
&=& - \int_{A}^{B} (P_2 - P_1) dx = {\Delta m^2 \over 2E} L \; .  
\eea
This computation seems not to use the
null condition $dx$ = $dt$, and it does not use the fact that, because the
velocities are not the same, the neutrinos with velocities $v_1$ and $v_2$
follow different world-lines. However, the null condition is in fact used 
when we replace both $\overline{BD_1}$ and $\overline{BD_2}$ by
$\overline{BN}$ (see Fig.\ref{figdv}). It is thus important to
remark that it is the null condition, not the same energy condition, that
accounts for the cancellation of the time phase.

What does it happen if the real world-lines (geodesic) of the neutrinos are taken
into account? In order to answer this question, it is important to notice that
the massive neutrinos $\nu_1$ and $\nu_2$ describe two different world lines 
from the source {\bf A} to the detector {\bf B}, given respectively by $AD_1$ and
$AD_2$ (see Fig.\ref{figdv}). Taking this into account, we obtain: 
\bea
\label{edp}
\Delta { \Phi} 
&=& \int_{A}^{D_2} (E_2 dt - P_2dx) - \int_{A}^{D_1} (E_1 dt - P_1 dx)
    \nonumber \\
&=& \int_{D_1}^{D_2} E dt - \int_{A}^{B} (P_2 - P_1) dx \nonumber \\ 
&=& {\Delta m^2 E \over 2 P_1 P_2} L +
    {\Delta m^2 \over 2 E} L \nonumber \\
& \approx & {\Delta m^2 \over 2 E} L + 
    {\Delta m^2 \over 2 E} L = 2 \Delta \Phi ^S \; .
\eea
If the arrival time-difference of the neutrinos is considered, therefore,
the time-phase is not canceled. As it is as large as the space-phase, the
real phase results twice the value yielded by the standard treatment. A
similar conclusion has also been obtained in Ref.~\cite{gro96,lip98}. The
computation of (\ref{edps}) and (\ref{edp}) indicates that it is the null 
condition, not the same energy condition, the responsible for the duplication
of the standard phase difference. 

\subsection{Reexamination of the Same-Momentum Prescription}

Following a procedure similar to that used in the case of the same-energy
prescription, we examine now the same-momentum prescription ($P_1 = P_2 = P$, 
$E_j^2 - P^2 = m_j^2$). We still suppose $m_2 > m_1$, which leads to $\gamma
_1 > \gamma _2$, $E_1 <  E_2$ and $v_1 > v_2$. These relations allow us to
use the same Minkowski diagram of Fig.\ref{figdv}. We find in this case
\bea
\label{pdp}
\Delta {\Phi}
&=& \int_{A}^{D_2} (E_2 dt - P_2 dx) - 
    \int_{A}^{D_1} (E_1 dt - P_1 dx) \nonumber \\
&=& \int_{B}^{D_2} E_2 dt - \int_{B}^{D_1} E_1 dt \nonumber \\
&=& {\Delta m^2 \over P} L \approx 2 \Phi^S \; .
\eea                
It is the same result of the case of the same-energy prescription. However,
a small difference exists. In order to show the influence of the same-momentum
and of the null-condition on the calculation of the phase difference, we
examine the phase computation of the same-momentum prescription of the
standard treatment, but using the null condition. This amounts to replace
$BD_2$ and $BD_1$ by the null time corresponding to $BN$. We find 
\bea
\label{pdps}
\Delta {\Phi}^S 
&=& \int_{A}^{B} (E_2 dt - P_2 dx) - \int_{A}^{B} (E_1 dt - P_1 dx)
    \nonumber \\
&=& \int_{B}^{N} E_2 dt - \int_{B}^{N} E_1 dt \nonumber \\
&=& \int_{B}^{N} (E_2 - E_1) dx = {\Delta m^2 \over 2P} L = \Phi^S \; .
\eea               
We see in this way that the same-momentum condition cancels the space phase
in both (\ref{pdp}) and (\ref{pdps}). However, the time-phase in (\ref{pdp})
is twice the value found in (\ref{pdps}).

A problem then arises: Why do both the same-energy and the same-momentum
prescriptions yield the same practical
result? A possible answer is that the
time-phase and the space-phase might be equivalent when the null condition is
used, as already concluded in Ref.~\cite{gro96}. Somehow, the null condition
implies in neglecting the arrival time-difference of the two neutrinos. If this
is correct, we should get $\Phi^S$ when computing the phase factor under the
assumption of the same-velocity prescription, as in this case no arrival
time-difference exists.

\subsection{On the Same-Velocity Prescription}

Although the same velocity descrition \c{tak98,leo99,tak00} is forbidden 
by the kinematical consideration \c{oku007}, 
here, we still explore in the Minkowski diagram how it 
arises the standard oscillation phase.   

Instead of supposing $E_1 = E_2$ or $P_1 = P_2$, let us suppose the
same-velocity prescription for the neutrinos motion~\cite{tak98,leo99,tak00}: $v_1$ =
$v_2 $ ($\gamma_1$ = $\gamma_2 $). 
We have in this case 
\be 
{m_2 \over m_1} = {E_2 \over E_1} = {P_2 \over P_1} \; ,
\label{veq1}
\ee
and
\be
{P_1 \over E_1} = {P_2 \over E_2}  \; .
\label{veq2}
\ee
The phase difference can be computed along the world line shown in
Fig.\ref{figsv}. The result is
\bea
\Delta {\Phi } 
&=& \int_{A}^{D} (E_2 dt - P_2 dx) - 
\int_{A}^{D} (E_1 dt - P_1 dx) \nonumber \\
&=& \int_{B}^{D} {(P_2 -  P_1) \over v_o}
\left({dx \over v_o} \right) - \int_{A}^{B}(P_2 - P_1) dx \nonumber \\
&=& {(P_2 - P_1) \over \gamma^2 v_o^2 } L =
    {(P_2^2 - P_1^2) \over \gamma^2 v_o^2 (P_2 + P_1 )} L \nonumber \\
&=& {\Delta m^2 \over (P_2 + P_1)} L \approx \Delta { \Phi}^S \; .
\eea 
Like the same-energy and the same-momentum prescriptions, the
same-velocity
prescription gives exactly the value of the standard treatment. We conclude
in this way that the confusion on the understanding of the neutrino phase
factor has its origin in the arrival-time difference.
 The null condition of the standard treatment includes the information of
the same speed of the two neutrinos. 

\section{geodesic phase and null phase in the curved spacetime}
As discussed in the flat spacetime, calculating  the phase along the 
geodesic  will produce a factor of 2, now we can also obtain 
this conclusion in the general curved spacetime, not only in the
Schwarzschild spacetime \cite{bha99}. 
The velocity of an extremely relativistic neutrino is nearly the speed of
light in the curved spacetime.  Although seemingly irrelevant to think
about the difference between the geodesic and the null, this
tiny
deviation becomes  important for the understanding of the neutrino
oscillation.  Motivated by this argument, we will compare the   
neutrino
phase when calculated along the geodesic and along the null-line. With
this,
we will be able to verify the factor of 2    when the null
is replaced by the geodesic. This study can be shown to remain valid in
the case of the flat spacetime.

Let $n^{\mu}$ and ${\stackrel{\circ}{n}}{}^{\mu}$ be the tangent vectors
to  
the geodesic and to the null-line, respectively, their difference
$\ep^{\mu}$ 
being a small quantity for the case of an extremely relativistic neutrino.
Here, we suppose that the two neutrinos, the massless and massive, start
their journey at the same initial spacetime position A, and their {\it
space}
routes are almost the same. But, there will be an arrival time-difference
at
the  detector position B. This means that their 4-dimensional spacetime
trajectories are not the same, and consequently the tangent vectors will  
present a small difference. Thus, we have
\be \label{nne}
n^{\nu} = {\stackrel{\circ}{n}}{}^{\nu} + \ep^{\nu}, \quad \mbox{or} \quad
P^{\nu} = {\stackrel{\circ}{P}}{}^{\nu} + m\ep^{\nu} \; ,
\ee 
where $P^{\nu} = m n^{\nu}$ (${\stackrel{\circ}{P}}{}^{\nu}= m
{\stackrel{\circ}{n}}{}^{\nu}$) is the 4-momentum along the geodesic
(null-line) with
\be
{\stackrel{\circ}{n}}{}^{\nu} = {dx^{\mu}\over d\lambda} =
{d{\stackrel{\circ}{x}}{}^{\nu}\over ds}
\ee
and
\be
n^{\mu} = {dx^{\mu}\over ds} \; .
\ee
In these expressions, $\lambda$ and $s$ are respectively affine parameters
along the null and the geodesic lines. These two tangent vectors satisfy
the
mass shell relations of the geodesic and the null-line:
\be\label{gn}
g_{\mu\nu} n^{\mu}n^{\nu} = 1
\ee 
and
\be\label{nn}
g_{\mu\nu} {\stackrel{\circ}{n}}{}^{\mu}{\stackrel{\circ}{n}}{}^{\nu} = 0
\; . 
\ee
Now, substituting (\ref{nne}) into (\ref{gn}), we obtain
\be \label{msg2}
g_{\mu\nu} ({\stackrel{\circ}{n}}{}^{\mu} + \ep^{\mu})
 ({\stackrel{\circ}{n}}{}^{\nu} + \ep^{\nu}) = 1 \; ,
\ee
or, by using (\ref{nn}),
\be \label{nne2}
2 g_{\mu\nu} {\stackrel{\circ}{n}}{}^{\mu} \ep^{\nu} +
{\cal O}(\ep^{2}) = 1 \; .
\ee
We can estimate the order of $\{n^{\mu}\}$ and $\{
{\stackrel{\circ}{n}}{}^{\mu} \}$ by noting that
$ n \ep \sim 1/2 $, which implies that $\ep \sim n^{-1} \sim {m\over E}$,
where $E = P^{o} \sim P^{i}$ {} $(i=1,2,3)$ for a relativistic neutrino.

The neutrino phase induced by the null condition, as in the standard
treatment, comes from the 4-momentum $P^{\nu}$ defined along the geodesic
line, and the tangent vector $\{ {\stackrel{\circ}{n}}{}^{\mu} \}$ to the
null-line~\cite{ful96}. We notice that, if the 4-momentum
${\stackrel{\circ}{P}}{}^{\nu}$ defined along the null-line was instead
used
to compute the null phase, we would obtain zero because of the null
condition.
Therefore, the  phase along the geodesic line (geodesic phase) and the
phase
along the null-line (null phase) can be written respectively
as~\cite{ful96,sto79,aud81,ana}
\be\label{nrpb}
\Phi({\rm geod})= \int m ds = \int
g_{\mu\nu}P^{\mu}n^{\nu}ds \; ,
\ee
and 
\be
\Phi({\rm null}) = \int g_{\mu\nu}P^{\mu}{\stackrel{\circ}{n}}{}^{\nu}ds
\; .
\ee 
Therefore, the difference between the geodesic phase and the null phase,
by using Eq.(\ref{nne2}), is
\bea \nn
\Phi({\rm geod})& -& \Phi({\rm null})= \int
g_{\mu\nu}P^{\mu}(n^{\nu} - {\stackrel{\circ}{n}}{}^{\nu}) ds\\\nn
&= &\int g_{\mu\nu}P^{\mu} \ep^{\nu} ds  =
 \int g_{\mu\nu}{\stackrel{\circ}{P}}{}^{\mu}\ep^{\nu} ds +
{\cal O}(\ep^{2})\\\nn
&=& {1\over2}\int m ds + {\cal O}(\ep^{2})
 =  {1\over2}\Phi({\rm geod}) + {\cal O}(\ep^{2}) \; ,
\eea
that is
\be
\Phi({\rm geod}) = 2 \Phi({\rm null}) + {\cal O}(\ep^{2}) \; .
\ee  
This conclusion, valid for a general curved spacetime, is similar to that
found in in a Schwarzschild~\cite{bha99} spacetime. Concerning
the Schwarzschild spacetime, Bhattacharya {\it et al}~\cite{bha99} have
the
following argument for the factor of 2. As the neutrino energy is fixed,
but the masses are different, if an interference is to be observed at the
same
final spacetime point B$(r_B,t_B)$, the relevant components of the wave
function could not both have started at the same initial spacetime point
 A$(r_A,t_A)$ in the semiclassical approximation. Instead, the lighter  
mass
(hence faster moving) component must either have started at the same time
from
a spatial location $r<r_A$, or (what is equivalent) started from the same
location $r_A$ at a later time $t_A + \Delta t$. Hence, there is already
an
initial phase difference between the two mass components due to this time
gap,
even before the transport from $r_A$ to $r_B$ which leads to the phase
$\Phi({\rm null})$, {\it i.e.}, the additional initial phase difference
may be 
taken into account~\cite{bha99}.

\section{discussions and conclusions}
As mentioned above, 
 the same-energy and the same-momentum prescriptions   
in calculating the mass neutrino 
 interference phase will obtain the same result if the arrival time 
difference is taken into account, which will be, mathematically,  equivalent to 
 the  phase of a   treatment by using the same-velocity  condition. 
However, we remark that  generally neither energy
nor momentum are equal in the factual physical process \c{dod99,giu01}, 
 the standard treatment of the 
 same-energy and  the same-momentum prescriptions of massive neutrino phases 
is just a mathematical simplification in some sense when applying 
the plane wave approximation. 
Then our analysis  of the  massive neutrino standard phases in the Minkowski diagram 
indicates that the same phase factor will be obtained if the two massive neutrinos follow the 
null line  or possess the same velocity. 
If the phase calculated 
along the particle world line, the realtive phase will produce a factor 
of 2 \cite{lip98}, the reason is that we despise the arrival time
difference 
of the two mass neutrinos, which results in a double counting effect. 
 This conclusion is correct in both flat spacetime and the curved
spacetime. 
Further, for the better understanding of the mass neutrino interference, 
 we should consider the neutrino mixing state as a  wave-packet \c{dod99,giu91}, 
or a
relativistic quantum ball. According to this scheme, two physical aspects 
must be considered. On one hand, there is the classical trajectory of the 
ball which is defined by giving the initial conditions in the classical
sense. This motion is relativistic, leading thus to the relativity of the
simultaneity as the velocities of the different mass neutrinos are not the
same. The classical trajectory, therefore, is defined in terms of macroscopic
quantities of the packet, the quantum average of quantities, in accordance 
with  
the correspondence principle of quantum mechanics.  In the classical sense,
the average mass and velocity of the packet ball define its orbit. The distance
from the source to the detector is usually measured accurately, which is a
classical measurement with a macroscopic precision. 
On the other hand, there is the quantum dynamics of the packet, which
is a two-state system like a neutral kaon or a B meson.
This fuzzy dynamics takes place in the microscopic scale $d$, the
characteristic dimension of the ball, which is small compared to the dimension
$L$ of the macroscopic trajectory.  This quantum dynamics, that is, the internal
evolution of the packet, is well described and understood in the particle physics
context.

As a final comment, we would like to argue that, if the individual
mass-eigenstate is supposed to have a well defined classical
velocity, its orbit would be well defined at any spacetime position. This,
however, is not in accordance with the quantum interference description, which
consequently would never occur. This means that the quantum fuzzy is necessary
for the wave-packet, a well defined orbit being valid only for the classical
(macroscopic) quantum average. According to this point of
view, two basic points concerns the description of the neutrino oscillation. For
the kinematics, we use the plane-wave description with a Lorentz
invariant phase. For the dynamics, we use the wave-packet and fuzzy quantum
path. We use, therefore, two concepts for the neutrinos, particle
and wave, classical and quantum. The neutrino is described in terms of a spinor
wave-packet, and the spacetime translation induces the spinor to precess in
the state vector-space. In other words, we take the mixing state as a classical
object propagating along the classical world-line with a well defined velocity ---
the group velocity. The individual mass-eigenstates propagate with a phase
velocity, and only the relative phase velocity will produce the realistic
interference phenomenon.

\section*{Acknowledgments}

Thanks are due to the scientific visiting  support by  ICTP, Trieste, Italy. 
A.B. is supported by  NRF of  South Africa.
 Discussions with A. Smirnov are highly appreciated. 
 The authors are grateful to the modification suggestions from the anonymous 
 referee.

\begin{figure}

\caption[fig1.eps]
{ 
For the case of same-velocity neutrinos, 
the Lorentz transformation makes the moving axis inclined to 
the null-line. The primed coordinate frame represents the neutrino
system moving with velocity $v$ in relation to the laboratory system
$(x, t)$.}
\label{figsv}
\end{figure}

\begin{figure}
\caption[fig2.eps]
{For the case of different-velocity neutrinos, different
Lorentz transformations make the corresponding coordinate system
differently inclined to the null-line. The faster the object the
closer they are of the null-line. The primed and double primed
coordinate frames represent respectively neutrinos
$\protect{\nu_1}$ and $\protect{\nu_2}$ with velocities $v_1$ and
$v_2$ relative to the laboratory system $(x, t)$.}
\label{figdv}
\end{figure}

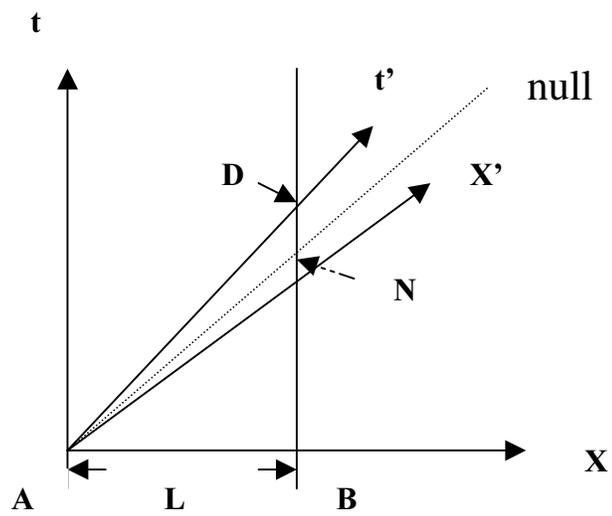
\end{document}